\documentclass[final,3p,times,twocolumn]{elsarticle}

\usepackage[utf8]{inputenc} 
\usepackage[T1]{fontenc}    
\usepackage{hyperref}       
\usepackage{url}            
\usepackage{booktabs}       
\usepackage{amsfonts}       
\usepackage{nicefrac}       
\usepackage{microtype}      
\usepackage{lipsum}
\usepackage{caption}
\usepackage{siunitx} 
\usepackage{graphicx}
\usepackage{amsmath}
\usepackage{amssymb}
\usepackage{multirow}
\graphicspath{{media/}}     %

\begin{document}

\begin{frontmatter}

\title{Haematocrit and shear rate modulate local cell-free layer thickness and platelet margination in blood flow along a sinusoidal wall}

\author{Eleonora Pero\fnref{label1}}
\author{Giovanna Tomaiuolo\corref{cor1}\fnref{label1,label2}}
\ead{g.tomaiuolo@unina.it}
\author{Stefano Guido\fnref{label1,label2}}
\author{Claire Denham\fnref{label3}}
\author{Timm Krüger\corref{cor2}\fnref{label3}}
\ead{timm.krueger@ed.ac.uk}

\affiliation[label1]{organization={Dipartimento di Ingegneria Chimica, dei Materiali e della Produzione Industriale, Università degli Studi di Napoli ``Federico II''},
             addressline={Piazzale Tecchio 80},
             city={Napoli},
             postcode={80125},
             state={Italy}}

\affiliation[label2]{organization={CEINGE Biotecnologie Avanzate ``Franco Salvatore''},
             addressline={via Gaetano Salvatore 486},
             city={Napoli},
             postcode={80131},
             state={Italy}}

\affiliation[label3]{organization={School of Engineering, Institute for Multiscale Thermofluids, University of Edinburgh},
             addressline={King's Buildings},
             city={Edinburgh},
             postcode={EH9 3FB},
             state={Scotland, UK}}

\cortext[cor1]{Corresponding author, address: Piazzale Tecchio 80, Napoli, 80125, Italy}

\cortext[cor2]{Corresponding author, address: James Clerk Maxwell Building, King's Buildings, Edinburgh, EH9 3FB, Scotland, UK}

\begin{abstract}
The geometry of blood vessels strongly affects hemostasis and thrombosis through red blood cell (RBC) dynamics and platelet margination. Growing platelet aggregates, in turn, reshape the local vessel wall topography, leading to a strongly coupled system. However, it is not well understood how surface heterogeneities alter local hemodynamics and platelet margination, thereby driving further aggregate growth.
This study investigates how hematocrit (\textit{Ht}) and shear rate affect RBC dynamics, cell-free layer (CFL) thickness, and platelet margination near a sinusoidal wall.
The sinusoidal wall, with crests and valleys aligned with the flow direction, serves as a model of the flow-aligned platelet aggregates observed in microfluidic experiments [Pero \textit{et al}., CRPS, 2024].
We perform three-dimensional immersed-boundary-lattice-Boltzmann simulations of particulate blood flow with deformable RBCs and nearly rigid spherical platelets.
Our results show that platelet margination is primarily governed by \textit{Ht} and is more pronounced in regions where the CFL thickness is similar to the platelet size.
At low \textit{Ht}, platelets preferentially accumulate at crests, promoting high-amplitude aggregate growth. Increasing \textit{Ht} leads to a more uniform platelet distribution along the surface, consistent with experimental observations.
The sinusoidal geometry generates a pronounced crest–valley wall shear rate gradient, suggesting that distinct shear-dependent adhesion pathways may dominate at different surface locations.
Our findings provide mechanistic insights into the morphological evolution of platelet aggregates and may ultimately inform targeted therapeutic strategies for thrombosis based on shear-sensitive drug-delivery.
\end{abstract}

\end{frontmatter}


\section{Introduction}

Platelet aggregation is the final stage of primary hemostasis.
Platelets are able to recognize regions of damaged endothelium, adhere, and recruit additional platelets to form aggregates that seal the injury and prevent bleeding.
Dysregulation of hemostatic processes and abnormal growth of blood clots may lead to vessel obstruction and thrombosis.
Hematocrit (\textit{Ht}) and wall shear rate ($\dot{\gamma}_{\text{w}}$) are key regulators of platelet transport toward the wall, adhesion, and aggregation.

The lateral migration of platelets toward the wall, termed \textit{margination}, produces a platelet concentration near the wall several times higher than in the bulk~\cite{Tangelder1982,Tilles1987,Aarts1988}.
This near-wall excess increases the probability of platelets adhering to the damaged endothelium~\cite{Yeh1994}.
The concentration peak usually lies within or near the cell-free layer (CFL) which is depleted of RBCs~\cite{Yeh1994, DApolito2015, DApolito2016}.
The CFL arises from two competing mechanisms: (i) hydrodynamic lift that pushes deformable RBCs away from the wall~\cite{Coupier2008}, and (ii) shear-induced diffusion (SID), caused by homogeneous RBC–RBC collisions that tend to disperse RBCs~\cite{Salame2025, Grandchamp2013}.
In the bulk, platelets undergo SID via collisions with RBCs, and follow random trajectories~\cite{Zhao2011}.
Near the edge of the CFL, these interactions become asymmetric, creating a drift bias that pushes platelets toward the wall~\cite{Fogelson2015, Yeh1994}.
Simulations indicate that the non-diffusive platelet transport near the CFL edge becomes important when RBCs are in the tank-treading regime (roughly $\textit{Ca} > 0.2$), which facilitates platelet trapping in the CFL~\cite{Krueger2015}.
Furthermore, in the tank-treading regime, platelet margination is relatively insensitive to the shear rate~\cite{Reasor2013, Mueller2014, Krueger2015}, but is primarily affected by \textit{Ht}~\cite{Spann2016} and confinement~\cite{Krueger2015}. CFL thickness regulates the degree of platelet margination, with particles more prone to margination when the CFL thickness becomes comparable to their size~\cite{Mueller2014, Mueller2016}.
Numerical simulations also show that platelet margination depends on particle shape and aspect ratio, with spherical platelets being more prone to margination than disc-shaped ones~\cite{Mueller2014}.
The biological pathway involved in platelet adhesion is determined by the wall shear rate.
In low-shear regions, platelets interact directly with the exposed extracellular matrix, predominantly via collagen-platelet $\alpha_2 \beta_1$ binding~\cite{Sixma1995}.
Longer residence times and reduced washout in these regions further promote bond formation and stabilization~\cite{Herbig2017}.
At shear rates above $\sim 500~\text{s}^{-1}$, platelet capture occurs mainly via bonding between platelet $\mathrm{GPIb}\alpha$ and the A1 domain of the von Willebrand factor (vWF) immobilized on the damaged surface~\cite{Jackson2007}.
At still higher shear ($\sim 1000~\text{s}^{-1}$ and above), hydrodynamic forces elongate vWF, increasing A1 exposure and facilitating GPIb$\alpha$-vWF interactions~\cite{Schneider2007}.
Under high shear, platelets frequently undergo pronounced translocation, rolling and sliding along the wall under GPIb$\alpha$-vWF bonds, traveling long distances before firm arrest.
These long surface excursions, driven by convection, promote aggregate growth in the flow direction and help explain the flow-aligned morphologies reported in microfluidic experiments~\cite{Colace2010, Pero2024}.
Consistent with this, Colace~\textit{et al.} observed that, at elevated shear ($500$--$2000~\text{s}^{-1}$), platelet deposits on collagen-coated surfaces elongate approximately twofold in the streamwise direction~\cite{Colace2010}.
Additionally, higher \textit{Ht} increases platelet margination and adhesion on collagen-coated surfaces~\cite{Spann2016, Walton2017}, resulting in higher and more uniform surface coverage and wider aggregates~\cite{Pero2024}.

The presence of mural platelet aggregates reshapes the local topography of the vessel wall.
Geometric heterogeneities can alter local hemodynamics, with consequences for near-wall platelet transport and adhesion.
Image-based simulations around platelet aggregates reveal pronounced spatial variations in the wall shear rate on the aggregate surfaces~\cite{Hao2023, Hao2024}, indicating that the evolving surface geometry feeds back on local flow.
Moreover, in the presence of surface heterogeneities~\cite{Zhang2023} or variations in the cross-section~\cite{Recktenwald2023}, RBC lift and the resulting CFL thickness vary along the wall.
However, most mechanistic studies on platelet margination focus on straight smooth walls, and the effect of a heterogeneous wall geometry on platelet margination remains poorly understood.

The present study addresses this gap using resolved three-dimensional simulations of blood flow near a sinusoidal wall.
The wavy geometry mimics the regular pattern arising from streamwise-elongated platelet deposits, observed in microfluidic experiments~\cite{Colace2010, Pero2024}.
The sinusoid amplitude and wavelength are set on the order of the RBC size, to maximize cell–wall interactions while remaining within experimentally observed length scales.
Numerical methods and the simulation details are explained in section~\ref{sec:methods} and section~\ref{sec:setup}, respectively.

This study aims to (i) investigate the influence of the vessel wall geometry on the local CFL profile, including correlations (Spearman, False Discovery Rate (FDR) 1\%) with RBC nematic ordering (section~\ref{sec:Res_CFL}); (ii) identify preferred sites for platelet capture along the sinusoidal surface and correlate them to the local CFL thickness (section~\ref{sec:Res_plt}); and to (iii) quantify the wall shear rate gradient between crest and valley, and interpret it in relation to the established shear-dependent regimes of platelet adhesion (section~\ref{sec:Res_WSR}).
The biochemical pathways of platelet adhesion are not explicitly modeled.
However, the mechanistic context for where different pathways are likely to dominate is provided by the wall shear rate (collagen/GPVI at lower shear~\cite{Sixma1995}; vWF-GPIb$\alpha$ capture and translocation at higher shear~\cite{Herbig2017, Jackson2007, Schneider2007}).
We systematically vary \textit{Ca} and tube \textit{Ht} to parse their roles in CFL thickness, platelet margination and wall shear rate along the crest and valley of the sinusoidal wall.
Together, these results help explain the morphological evolution of platelet aggregates observed in microfluidic experiments~\cite{Colace2010, Pero2024}.
Finally, the identification of margination hotspots, coupled with the probable adhesion pathway, may offer a rational basis to inform shear-sensitive antithrombotic strategies~\cite{ZeibiShirejini2023, Molloy2017} and the design of microfluidic assays.

\section{Numerical methods}
\label{sec:methods}

\begin{figure*}[t]
 \centering
 \includegraphics[width=1\textwidth]{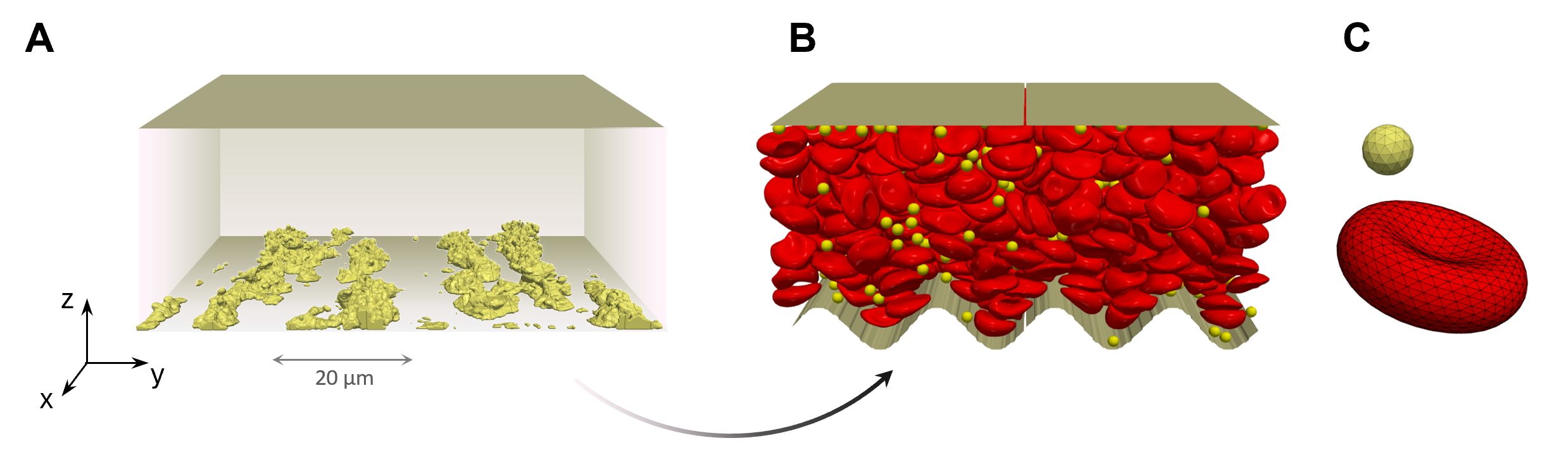}
 \caption{Simulation of blood flow in a straight channel with a sinusoidal bottom wall, resembling mural platelet aggregates.
 \textbf{(A)} 3D reconstruction from microfluidic experiments of platelet deposits on a collagen-coated surface, under physiological \textit{Ht} ($\approx 0.45$) and wall shear rate of $500~\text{s}^{-1}$. The aggregates (shown in yellow) are aligned with the flow direction ($x$-axis). Experimental details are reported elsewhere~\cite{Pero2024}.
 \textbf{(B)} Numerical simulation of blood flow in a straight channel with a sinusoidal bottom surface, mimicking the geometry of platelet aggregates observed in the microfluidic experiments. RBCs are represented in red, platelets in yellow.
 \textbf{(C)} Unstructured surface meshes employed in the numerical model. Each RBC is discretized with 1280 triangular elements, while platelets, modeled as nearly rigid spheres, are represented with 180 triangular elements.
 }
 \label{Figure_1}
\end{figure*}

The present research concerns with blood flow in a straight channel with a flat top wall and a sinusoidal bottom wall.
The sinusoidal surface serves as a simplified model of elongated platelet aggregates commonly observed in microfluidic experiments, as shown in Fig.~\ref{Figure_1}~\cite{Colace2010, Pero2024}, while a straight upper wall provides a control.
Simulations were performed using the in-house code \textit{BioFM} which is designed to simulate blood flow in the microvasculature, resolved on a cellular scale.
The numerical model uses the lattice-Boltzmann method as Navier–Stokes solver (section~\ref{sec:LBM}) and the immersed-boundary method to couple the fluid and cell dynamics (section~\ref{sec:IBM}).
RBC membranes are modeled with a finite-element approach (section~\ref{sec:FEM}).
Boundary conditions are described in section~\ref{sec:boundaryconditions}.
Further information and validation tests are reported elsewhere \cite{Krueger2011, Krueger2012, Krueger2013, Krueger2014}.

\subsection{Lattice-Boltzmann method}
\label{sec:LBM}

In this work, the standard lattice-Boltzmann method (LBM) with the D3Q19 lattice~\cite{Qian1992} and the Bhatnagar–Gross–Krook (BGK) collision operator~\cite{Bhatnagar1954} is employed to solve the Navier–Stokes equations (NSE)~\cite{Krueger2017}, with $\Delta x$ and $\Delta t$ being the lattice resolution and the time step, respectively.
Guo's forcing scheme~\cite{Guo2002} is used to include immersed-boundary forces and the force driving the flow along the channel.

\subsection{Immersed boundary method}
\label{sec:IBM}

The immersed boundary method (IBM)~\cite{Peskin1972, Peskin2002} couples the fluid flow on the Eulerian lattice and the blood cells described as Lagrangian particles.
IBM uses interpolations to enforce the no-slip condition at the surface of immersed particles.
Conversely, the particles exert local force on the surrounding fluid, resembling momentum exchange between particles and fluid.
These forces are first spread to the Eulerian lattice before being applied to the fluid through the employed forcing scheme of the LBM.
We use a 2-point trilinear interpolation stencil~\cite{Peskin2002, Krueger2015}.

\subsection{Red blood cell elasticity model}
\label{sec:FEM}

RBCs and platelets are treated as closed membranes discretized by $N_f$ flat triangular elements.
The generation of the particle mesh is based on the icosahedron-refinement procedure~\cite{RAMANUJAN1998, Krueger2011}.
The undeformed stress-free shape of the RBC is illustrated in Fig.~\ref{Figure_1}~C, resembling a biconcave disc with a radius of about \SI{4}{\micro\meter}.
In the present work, each RBC mesh consists of 1280 elements.
Platelets are assumed to be non-activated and modeled as nearly rigid spheres (Fig.~\ref{Figure_1}~C) with a radius of \SI{1}{\micro\meter}.
Each platelet mesh comprises 180 elements.

Under physiological conditions, RBCs are deformable while maintaining a nearly constant surface area $A^{(0)}$ and volume $V^{(0)}$~\cite{Evans1972}.
RBC membranes are characterized by two major elastic contributions: in-plane strain and bending resistance~\cite{Skalak1973, evans1974}.
In the present work, the RBC membrane is modeled as a hyperelastic continuum with four elastic energy contributions: (i) local in-plane forces, caused by the resistance to shear and dilation; (ii) local bending energy, giving rise to forces normal to the membrane; penalty energy terms to maintain constant (iii) surface and (iv) volume.
The in-plane energy density is described by Skalak’s constitutive law~\cite{Skalak1973}:
\begin{equation}
 E_S = \int \mathrm{d}A \left( \frac{\kappa_{\text{S}}}{12}\left(I_1^2 + 2I_1 - 2I_2\right) + \frac{\kappa_\alpha}{12} I_2^2 \right)
\end{equation}
where $\kappa_{\text{S}}$ and $\kappa_\alpha$ control the strength of the membrane 
response to local shear deformation and dilation, respectively.
For healthy RBCs, the values are $\kappa_{\text{S}} = \SI{5.3}{\micro\newton\per\meter}$ and $\kappa_\alpha = \SI{0.5}{\newton\per\meter}$~\cite{Krueger2015}.
The two strain invariants $I_1$ and $I_2$ describe the in-plane deformation and represent the local shear deformation and area dilation, respectively~\cite{Krueger2012}.
The bending energy~\cite{Krueger2015} is described by
\begin{equation}
 E_B = \frac{\sqrt{3}\,k_B}{2} \sum_{\langle i,j \rangle} \left( \theta_{ij} - \theta^{(0)}_{ij} \right)^{2}
 \label{bendingenergy}
\end{equation}
with the RBC bending resistance $\kappa_{\text{B}} = 2 \cdot 10^{-19}\,\si{\newton\meter}$, $\theta_{ij}$ the angle between the normal vectors $\hat{n}_i$ and $\hat{n}_j$ of two neighboring elements, and $\theta^{(0)}_{ij}$ the corresponding angle in the undeformed state.

In order to maintain nearly constant RBC surface area~\cite{evans1980mechanics, seifert1997configurations}  and volume~\cite{seifert1997configurations}, penalty energy contributions are introduced:
\begin{align}
 E_A &= \frac{\kappa_A}{2} \, \frac{\left( A - A^{(0)} \right)^2}{A^{(0)}}, \label{conssurface} \\
 E_V &= \frac{\kappa_V}{2} \, \frac{\left( V - V^{(0)} \right)^2}{V^{(0)}} \label{consvolume}
\end{align}
where $A^{(0)}$ is the undeformed surface area and $A$ is the current surface area.
For a healthy RBC, $A^{(0)} \approx \SI{140}{\micro\meter\squared}$.
The magnitude of the surface energy is controlled by the modulus $\kappa_A$, which is chosen sufficiently large to keep the surface area nearly constant.
A similar approach is used for the RBC volume, with $V$ and $V^{(0)} \approx \SI{100}{\micro\meter\cubed}$ being the current and undeformed RBC volume, respectively.

The principle of virtual work is used to calculate the local membrane forces from the energy contributions~\cite{Krueger2011}.

\subsection{Boundary conditions}
\label{sec:boundaryconditions}

The simulations are aimed at reproducing experimental conditions in straight microfluidic channels, typically of the order of centimeters in length ($l_x$), millimeters in width ($l_y$), and tens of micrometers in height ($l_z$).
Since a direct simulation of such large geometries at micrometer resolution would be computationally prohibitive, only a representative unit domain of reduced size ($L_x \times L_y \times L_z$ with $L_x \ll l_x$ and $L_y \ll l_y$) is considered and periodic boundary conditions~\cite{Krueger2017} for the fluid and blood cells are applied along the flow direction ($x$) and the transverse direction ($y$).
The flow is driven by a constant body force that is equivalent to a pressure gradient $p'$ along the flow direction, resulting in the desired average channel velocity $\bar{u}_0$.

The half-way bounce-back boundary condition is imposed at the sinusoidal and the straight walls to enforce the no-slip condition at these stationary walls~\cite{Krueger2017}.
In this scheme, populations that hit the rigid wall during propagation are bounced back to their original lattice site, leading to the desired no-slip condition.

\section{Simulation setup and system parameters}
\label{sec:setup}

After introducing the relevant geometrical parameters of the simulated domain in section \ref{sec:GP}, the choice of simulation parameters is briefly described in section \ref{sec:SP}.
Table~\ref{tab:par} provides an overview of all relevant parameters.

\begin{table}[tb]
 \centering
 \caption{Overview of simulation parameters}
 \label{tab:par}
 \scriptsize
 \begin{tabular}{lll} \hline\hline
  \textbf{Parameter} & \textbf{Symbol} & \textbf{Value} \\
  \hline
  Channel length           & $L_x$     & \SI{40}{\micro\meter} \\
  Channel width            & $L_y$     & \SI{40}{\micro\meter} \\
  Channel height           & $L_z$     & \SI{40}{\micro\meter} \\
  RBC radius               & $r_{\text{RBC}}$ & \SI{4}{\micro\meter} \\
  RBC confinement          & $\chi = 2 r_{\text{RBC}} / L_z$    & 0.2 \\
  Platelet radius          & $r_{\text{plt}}$& \SI{1}{\micro\meter} \\
  Sinusoidal amplitude     & $A_{\text{s}}$     & \SI{4}{\micro\meter} \\
  Reduced amplitude        & $\alpha = A / r_{\text{RBC}}$  & 1 \\
  Sinusoidal wavelength    & $\lambda_{\text{s}}$ & \SI{20}{\micro\meter} \\
  Tube hematocrit          & $Ht$      & Variable \\
  RBC count                & $N_{\text{RBC}}$ & Variable \\
  Platelet count           & $N_{\text{plt}}$& $N_{\text{RBC}}/2$ \\
  Fluid density            & $\rho$    & \SI{1000}{\kilogram\per\meter\cubed} \\
  Viscosity                & $\eta$ & \SI{1}{\milli\pascal\second} \\
  RBC shear elasticity     & $\kappa_{\text{S}}$ & \SI{5.3}{\micro\newton\per\meter} \\
  RBC bending elasticity   & $\kappa_{\text{B}}$ & $2 \cdot 10^{-19}\,\si{\newton\meter}$ \\
  Reduced bending modulus  & $\kappa_{\text{B}} / (\kappa_{\text{S}} r^2)$ & 1/424 \\
  Lattice resolution       & $\Delta x$ & \SI{0.33}{\micro\meter} \\
  Time step                & $\Delta t$ & Variable \\
  Centre velocity (no cells) & $\hat{u}_0$ & $0.05\,\Delta x/\Delta t$ \\
  Average velocity (no cells) & $\bar{u}_0$& $\hat{u}_0/2$ \\
  Straight wall shear rate & $\dot{\gamma}_{\text{w}}$ & $4\hat{u}_0/H_{\text{eq}}$ \\
  Pressure gradient        & $p'$       & $6.67 \cdot 10^{-6}$ s.u. \\
  Capillary number         & \textit{Ca}       & Variable \\
  Number of timesteps      & $N_t$      & $1.5 \cdot 10^5$ \\
  Advection time           & $t_{\text{adv}}$ & $735\,\Delta t$ \\
  \hline
 \end{tabular}
\end{table}

\begin{table}[tb]
 \centering
 \caption{Studied values for Ca}
 \label{tab:simparams}
\scriptsize
\renewcommand{\arraystretch}{1.2}
\setlength{\tabcolsep}{12pt}
 \begin{tabular}{ccc}
    \hline \hline
    \textit{Ca} & $\kappa_{\text{S}}$ (s.u.) & $\kappa_{\text{B}}$ (s.u.) \\
    \hline
    0.1 & 0.0400  & 0.00113  \\
    0.2 & 0.0200  & 0.000566 \\
    0.4 & 0.0100  & 0.000283 \\
    0.6 & 0.00667 & 0.000189 \\
    \hline
    \end{tabular}
    \end{table}

\subsection{Geometrical parameters}
\label{sec:GP}

The simulated domain consists of a straight planar channel with a sinusoidal bottom wall.
The sinusoidal surface resembles the elongated platelet aggregates commonly observed in microfluidic experiments (Fig.~\ref{Figure_1}). 
The geometric parameters of this surface are selected to approximate the geometry observed in experiments~\cite{Pero2024}.
The amplitude ($A_{\text{s}}$) of the sinusoidal surface is defined relative to the radius of the RBC ($\alpha = A_{\text{s}} / r_{\text{RBC}}$) with $\alpha = 1$ in the present study.
The wavelength ($\lambda_{\text{s}}$) of the sinusoidal wall was chosen as $\lambda_{\text{s}} = \SI{20}{\micro\meter}$, making the average curvature radius of the wall similar to the RBC radius.

The channel height was chosen as $L_z = \SI{40}{\micro\meter}$, leading to an RBC confinement of $\chi = {r_{\text{RBC}}}/{L_z} = 0.1$.
This channel height is sufficiently large to prevent platelet margination at the top wall to be influenced by the presence of the sinusoidal bottom wall.
$L_x$ and $L_y$ were set to \SI{40}{\micro\meter}, which is ten times the RBC radius, ensuring that effects of periodic images are negligible.

\subsection{System parameters}
\label{sec:SP}

All relevant simulation parameters are collected in Table~\ref{tab:par}. Simulations were performed at \textit{Ht} values of 0.33, 0.44 and 0.48, corresponding to a number of platelets ($N_{\text{plt}}$) of 106, 142 and 153, respectively. The values of \textit{Ca} and the corresponding values of $\kappa_{\text{S}}$ and $\kappa_{\text{B}}$ are reported in 
Table~\ref{tab:simparams}. 

A pressure gradient $p'$, mimicked by a constant and homogeneous force density, drives the flow along the channel in the positive $x$-direction.
The value of $p'$ is chosen in such a way that the peak velocity $\hat{u}_0$ in the absence of particles takes a desired value.
We keep the numerical value of $\hat{u}_0$ sufficiently small to avoid compressibility effects and achieve stable simulations.
Note that the actual peak velocity is smaller (typically by a factor of $\approx 2$) due to the presence of cells.

The deformation of an RBC with radius $r_{\text{RBC}}$ and shear elasticity $\kappa_{\text{S}}$ in a simple shear flow with viscosity $\eta$ and shear rate $\dot{\gamma}$ is characterised by the capillary number $\textit{Ca} = \eta  \dot{\gamma} r_{\text{RBC}} / \kappa_{\text{S}}$.
In pressure-driven flow, the shear rate is not constant and it is common to denote the scale of the shear rate by the wall shear rate $\dot{\gamma}_{\text{w}}$.
In this study, we use the wall shear rate at the straight upper wall to define \textit{Ca}.

Periodic boundary conditions along the $y$-axis eliminate lateral wall effects.
Additionally, since $L_z$ is substantially larger than the sinusoidal amplitude ($L_z / A_{\text{s}} \approx 10$) and the sinusoidal perturbation decays rapidly with distance ($\propto \exp (-2\pi L_z/\lambda) \approx 3 \cdot 10^{-6}$)~\cite{lauga2003}, the flow field at the straight upper wall can be approximated by that in a planar Poiseuille flow with an effective channel height.
In the presence of a flat lower wall, the velocity profile would be symmetric, with the maximum velocity in the middle between both walls.
However, the presence of the sinusoidal bottom wall shifts the position of the peak velocity upward.
The location of the velocity peak is extracted from the simulation data and found at a distance $d_{\text{top}} \approx \SI{17}{\micro\meter}$ from the upper wall.
The upper region of the channel can, therefore, be considered a ``half-channel'', bounded by the flat upper wall (no slip) and a virtual symmetry plane at the location of the peak velocity.
In this configuration, the cell-free velocity profile in the upper region of the channel can be assumed parabolic with a wall shear rate $\dot{\gamma}_{\text{w}} = p' d_{\text{top}} / \eta$ and the capillary number defined as
\begin{equation}
 \textit{Ca} = \frac{\eta \dot{\gamma}_{\text{w}} r_{\text{RBC}}}{\kappa_{\text{S}}} = \frac{p' d_{\text{top}} r_{\text{RBC}}}{\kappa_{\text{S}}}.
\end{equation}

To find a dimensionless time for the advection of the suspension, one may define the advection time scale as
\begin{equation}
 t_{\text{adv}} = \frac{2 r_{\text{RBC}}}{\bar{u}_0}
\end{equation}
with $\bar{u}_0$ the average fluid velocity in the absence of particles.
Each simulation runs for $N_t = 1.5 \cdot 10^{5}$ time steps, corresponding to around $200~t_{\text{adv}}$, ensuring a statistical steady state for the RBCs and a sufficiently long window to observe platelet margination.

The number of platelets in our simulations is $N_{\text{plt}} = N_{\text{RBC}}/2$, which is 7--8 times larger than observed under physiological conditions.
This choice gives a better platelet statistic while not changing the rheology of the suspension significantly.
At the beginning of a simulation, RBCs and platelets are initialised with random positions and orientations~\cite{Krueger2017, Krueger2015}.
To improve statistical reliability, all results are obtained by averaging over at least two independent simulations per condition, each with different initial particle distributions.

\subsection{Cell-free layer definition}
The CFL identifies the near-wall region depleted of RBCs. In this study, the CFL is defined as the region where the local RBC volume fraction is below 50\% of the tube $Ht$. 

The RBC volume fraction is computed and then averaged along the flow direction $x$ and over the final quarter of the simulation time ($100-200~t_{\mathrm{adv}}$). The CFL profile is identified as the locations $(y,z)$ at which the RBC local volume fraction reaches the $0.5~Ht$ threshold.

\subsection{Characterisation of particle deformation and orientation}

The near-wall deformation and ordering of RBCs are evaluated using the RBCs' inertia ellipsoid.
A particle's inertia ellipsoid is the (unique) ellipsoid with the same density and inertia tensor $\mathbf{T}$ as that of the particle.
By diagonalising $\mathbf{T}$ one obtains its ordered eigenvalues $T_1 \le T_2 \le T_3$ and the associated orthonormal eigenvectors.
From $(T_1, T_2, T_3)$, one can infer the semi-axes $a \ge b \ge c$ of the inertia ellipsoid~\cite{Krueger2012}.
Assuming an oblate particle, the eigenvector associated with the largest eigenvalue $T_3$ is aligned with the shortest semi-axis $c$ which is perpendicular to the $ab$-plane and thus defines the principal orientation of the particle in space.

For such an oblate particle, the time-dependent in-plane deformation index is defined as
\begin{equation}
 D(t) = \frac{a(t)-b(t)}{a(t)+b(t)},
\end{equation}
with $D = 0$ for an undeformed shape (if the oblate has rotational symmetry about the $c$-axis) and $D \approx 1$ for highly elongated shapes.

The orientational order of the RBCs is characterised by the nematic order tensor
\begin{equation}
 \mathbf{Q} = \frac{1}{2N} \sum_{i=1}^{N} \left(3\,\mathbf{e}_i\otimes\mathbf{e}_i - \mathbf{I} \right),
\end{equation}
where $\mathbf{e}_i$ is the unit orientation vector (normal to the $ab$-plane) of the $i$-th RBC and $\mathbf{I}$ is the identity tensor~\cite{dierking2003textures, tsige1999nematic, Krueger2012}.
The largest eigenvalue of $\mathbf{Q}$ is taken as the nematic order parameter $Q$: $Q \approx 1$ indicates strong alignment and $Q \approx 0$ suggests random orientations.

\section{Results and Discussion}
\label{sec:results}

The results and discussion are presented in three parts.
Section~\ref{sec:Res_CFL} studies the influence of the vessel geometry on the local CFL thickness, including correlations with RBC nematic ordering.
Section~\ref{sec:Res_plt} identifies platelet margination hotspots along the sinusoidal surface and correlates them with the local CFL thickness.
Section~\ref{sec:Res_WSR} quantifies the wall shear rate gradients between the crest and the valley and interprets them in relation to the established shear-dependent regimes of platelet adhesion.

\subsection{Cell-free layer and red blood cell ordering along the sinusoidal surface}
\label{sec:Res_CFL}

\begin{figure*}[tp]
 \centering
 \includegraphics[width=0.8\textwidth]{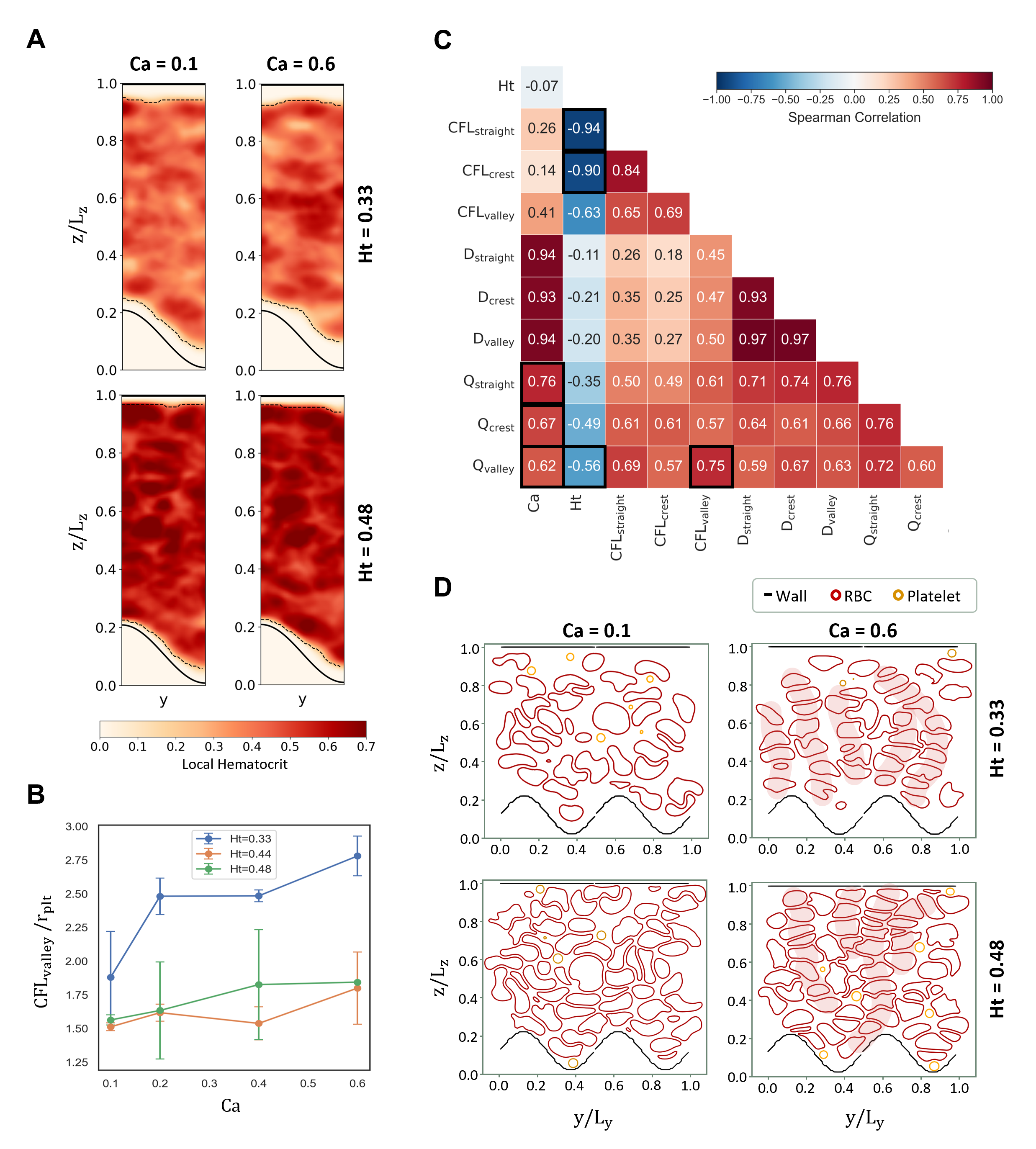}
 \caption [Effect of \textit{Ca} and tube \textit{Ht} on local \textit{Ht} and correlations between CFL, RBC deformation and ordering]{Effect of \textit{Ca} and tube \textit{Ht} on local \textit{Ht} and correlations between CFL, RBC deformation and ordering. \textbf{(A)} Average local \textit{Ht} (color map) versus cross-stream coordinate $y$ and normalised vertical coordinate $z/L_z$ ($z$ from bottom to top wall; $L_z$ total channel height) for two capillary numbers ($Ca = 0.1, 0.6$) and two \textit{Ht} values ($\textit{Ht} = 0.33, 0.48$). Data are averaged along the flow direction $x$, over the final quarter of the simulation time ($150-200~t_{\text{adv}}$), and then over equivalent positions within half of the sinusoidal wavelength of the lower wall, yielding a representative half-wave pattern. Solid line: wall profile. Dashed line: cell-free layer (CFL) profile, where local \textit{Ht} reaches 50\% of the tube \textit{Ht}. Color bar: average local \textit{Ht}. \textbf{(B)} CFL thickness at the valley ($\text{CFL}_{\text{valley}}$) normalised by the platelet radius ($r_{\text{plt}}$) versus \textit{Ca} for three values of \textit{Ht} ($\textit{Ht} = 0.33$, blue; $0.44$, orange; $0.48$, green). Data represent means and standard deviations from at least two simulations per condition with different particle initialisations. \textbf{(C)} Spearman correlation matrix between \textit{Ca}, \textit{Ht}, CFL thickness, RBC deformation ($D$) computed from the axes of the RBC inertia ellipsoids ($D=0$: undeformed; $D=1$: highly deformed), and the nematic order parameter ($Q$), which quantifies the alignment of RBC inertia ellipsoids ($Q=0$: random orientation; $Q=1$: perfect alignment). CFL, $D$ and $Q$ are displayed for the straight upper wall and for the crest and valley of the wavy lower wall. Color bar: Spearman correlation coefficient ($\rho$), with positive (red) and negative (blue) values indicating direct and inverse relationships, respectively. \textbf{(D)} Representative instantaneous cross-sections of the suspension after at least $100~t_{\text{adv}}$ for two \textit{Ca} values ($Ca = 0.1, 0.6$) and two \textit{Ht} values ($\textit{Ht} = 0.33, 0.48$). RBC contours in red, platelets in yellow, walls in black. Axes: $y/L_y$ (cross-stream) and $z/L_z$ (wall-normal). The red shaded area serves as a visual guide to highlight regions of ordered and packed RBCs.}
 \label{Figure_2}
\end{figure*}

The CFL along the sinusoidal wall is spatially heterogeneous, with a pronounced thickening in the valley, especially at high \textit{Ca} and low \textit{Ht} (Fig.~\ref{Figure_2}~A,B).
In contrast, the CFL at the crest and along the upper straight wall shows no significant correlation with \textit{Ca}, but a marked reduction with increasing \textit{Ht} (crest: $\rho = -0.90$, $p = 3.44 \times 10^{-9}$; straight wall: $\rho = -0.93$, $p = 6.24 \times 10^{-11}$; Fig.~\ref{Figure_2}~C), a trend that is also apparent in Fig.~\ref{Figure_2}~A.

These observations are consistent with previous studies reporting that CFL thickness is largely insensitive to \textit{Ca}~\cite{Spann2016}, reaching a plateau for $Ca \gtrsim 0.2$~\cite{Krueger2015}.
However, in complex geometries, such as corners~\cite{Rashidi2025, Rashidi2023}, cavities~\cite{Zhang2023} and stagnation regions~\cite{Wu2020}, increasing \textit{Ca} (or equivalently, RBC deformability) enhances RBC hydrodynamic lift and results in local CFL expansion.
At high \textit{Ht}, this lift effect is counterbalanced by stronger shear-induced dispersion, which redistributes RBCs across the vessel cross-section and reduces the CFL thickness~\cite{Zhang2009, Czaja2020}.

Fig.~\ref{Figure_2}~D highlights the emergence of ordered, packed RBCs in the channel core as \textit{Ca} increases from 0.1 to 0.6.
Accordingly, the nematic order parameter ($Q$) correlates positively with \textit{Ca} at the straight wall ($\rho = 0.76$, $p = 2.38 \times 10^{-5}$), crest ($\rho = 0.67$, $p = 4.3 \times 10^{-4}$), and valley ($\rho = 0.62$, $p = 1.7 \times 10^{-3}$).
In the valley, $Q_{\text{valley}}$ correlates negatively with \textit{Ht} ($\rho = -0.56$, $p = 5.6 \times 10^{-3}$; Fig.~\ref{Figure_2}~C).
This observation suggests that, at higher Ht, enhanced shear-induced dispersion and overcrowding force RBCs to occupy the entire cross-sectional volume and conform to the wavy geometry, rather than aligning with bulk cell orientation, as illustrated in Fig.~\ref{Figure_2}~D.

Importantly, $Q_{\text{valley}}$ correlates strongly with CFL thickness ($\rho = 0.75$, $p = 4.06 \times 10^{-5}$; Fig.~\ref{Figure_2}~C), supporting the hypothesis that RBC self-organisation into ordered, packed layers 
at high \textit{Ca} facilitates RBC hydrodynamic lift in the valley at low \textit{Ht}, thereby thickening the CFL.

Previous studies have shown that, under wall-confined shear, deformable RBCs can self-organise into crystal-like ordered structures, whereas rigid cells remain disordered~\cite{Shen2018}.
This ordering is driven by the combined effects of (i) hydrodynamic wall repulsion that pushes deformable cells towards the channel centre and (ii) intercellular hydrodynamic interactions~\cite{Shen2018}.
Moreover, Hariprasad~\textit{et al}.~\cite{Hariprasad2014} found that the orientation adopted by RBCs in the tank-treading regime ($Ca \gtrsim 0.2$~\cite{Krueger2015}), with a positive inclination of their long axis relative to the wall, strongly contributes to the net RBC hydrodynamic lift.
Together, these mechanisms support the interpretation that RBC self-ordering facilitates local CFL expansion in confined regions.

\subsection{Preferential sites of platelet capture along the sinusoidal surface}
\label{sec:Res_plt}

Identifying preferential adhesion sites along the wavy surface is fundamental for understanding the mechanisms underlying the temporal evolution of platelet aggregates.
The ability of a platelet to adhere to the vessel wall is primarily governed by its distance from the boundary~\cite{RUGGERI2009, Qi2019}.
In our model, biological adhesion pathways are not simulated; instead, we introduce a geometric criterion to estimate whether platelets (i) approach the wall sufficiently closely and (ii) are retained within the near-wall region, thereby providing favourable conditions for adhesion.

\begin{figure*}[tp]
 \centering
 \includegraphics[width=0.9\textwidth]{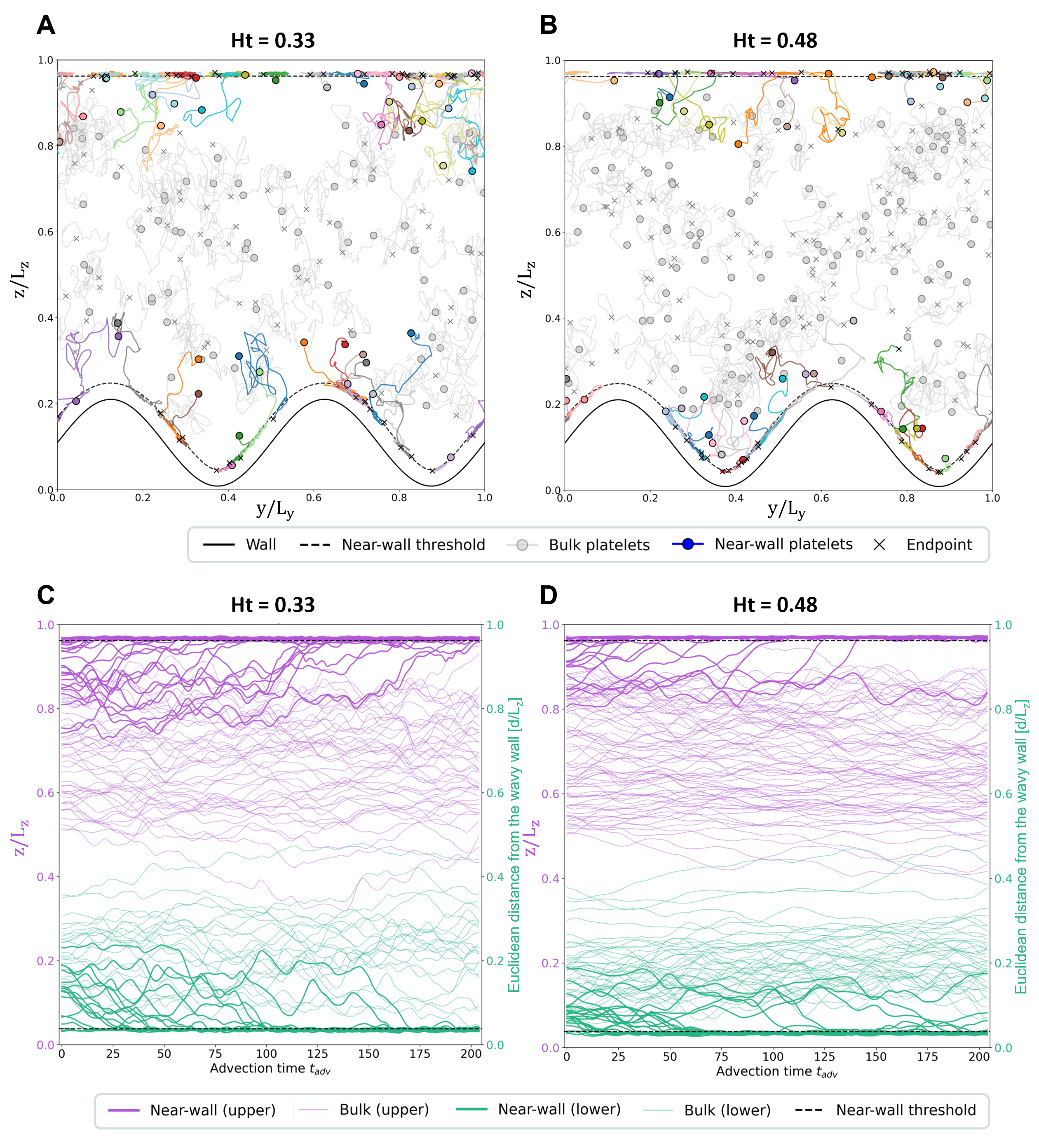}
 \caption {Analysis of platelet trajectories over $200~t_\text{adv}$. \textbf{(A--B)} Representative platelet trajectories in the $z$--$y$ plane over $200~t_\text{adv}$ at $\textit{Ca} = 0.4$. The solid lines denote the channel walls (straight upper wall and wavy lower wall), while the dashed line indicates the platelet capture distance ($1.5~r_{\text{plt}} = 1.5~\si{\micro\meter}$ from the wall). Coloured circles mark the initial positions of platelets that reach the capture distance at least once. Different colours are used to distinguish individual platelet trajectories for clarity. Grey circles and trajectories correspond to platelets that never reach the capture distance. Black crosses denote the endpoints of trajectories. A total of $n = 106$ platelets is tracked for $\textit{Ht} = 0.33$ (left), and $n = 153$ for $\textit{Ht} = 0.48$ (right). \textbf{(C--D)} Platelet-wall distance as a function of time, expressed in units of $t_\text{adv}$, for $\textit{Ca} = 0.4$, and $\textit{Ht} = 0.33$ (left) or $\textit{Ht} = 0.48$ (right). Purple lines represent platelets initialised in the upper half of the channel ($z/L_z > 0.5$). They refer to the left vertical axis, which reports the normalised vertical coordinate ($z/L_z$) and indicates the distance from the flat upper wall. Green lines represent platelets initialised in the lower half of the channel ($z/L_z \leq 0.5$) and refer to the right vertical axis (green), which indicates the Euclidean distance from the wavy lower wall (\textit{i.e.}, the minimum distance between the platelet and the wall). Bold trajectories highlight platelets that reach the capture distance (dashed lines) at least once.}
 \label{Figure_3}
\end{figure*}

\begin{figure*}[htb]
 \centering
 \includegraphics[width=1\textwidth]{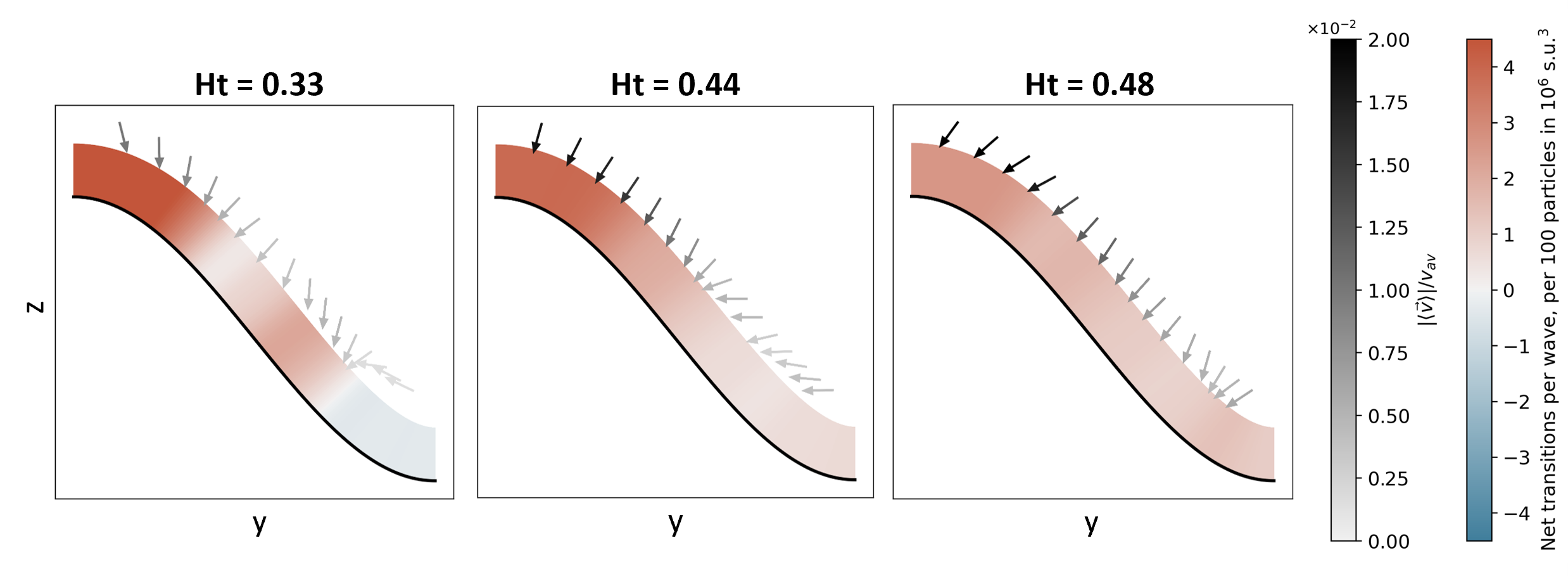}
 \caption{Spatial distribution of platelet capture along the sinusoidal surface. Heatmaps show the local probability of platelet capture, quantified as the net number of platelet transitions across the capture distance ($1.5~r_{\text{plt}}$ from the wall), over $\sim 200 t_\text{adv}$. Values are normalised to a reference density of 100 platelets in a simulated volume of $\sim 5.7 \times 10^{4}\,\mu\mathrm{m}^{3}$. Results are shown for three \textit{Ht} values ($\text{Ht} = 0.33, 0.44, 0.48$). Red regions denote the tendency of platelets to enter and accumulate near the wall, while blue regions indicate their tendency to exit toward the bulk. Arrows represent the mean platelet velocity within a $1.5$--$2.5~r_{\text{plt}}$ band from the wall, normalised by the average speed along the channel (greyscale, $\lvert \langle v \rangle \rvert / v_{\text{av}}$). Data are averaged across simulations with varying \textit{Ca} and initial particle distributions, and further averaged over equivalent positions within half a sinusoidal wavelength to obtain a representative half-wave pattern.}
 \label{Figure_4}
\end{figure*}

Specifically, the \textit{probability of platelet capture} is defined as the likelihood that the platelet center of mass reaches and remains within a capture region. It therefore represents near-wall retention, which is a prerequisite for stable platelet adhesion. Here, the capture region is defined as a distance from the wall of $1.5~r_{\text{plt}}$ ($1.5~\si{\micro\meter}$). The threshold accounts for the platelet radius ($r_{\text{plt}} = 1~\si{\micro\meter}$) and for an additional tolerance associated with a depletion layer that prevents particles from directly contacting the simulation boundary.
The chosen value is consistent with previous simulations where the capture distance was defined as the sum of the platelet radius and the length of von Willebrand factor (vWF)~\cite{Qi2019}, which varies with shear rate and has been experimentally estimated in the range of $\sim150$--$400\,\si{\nano\meter}$~\cite{Springer2014}.

The analysis of platelet trajectory is, therefore, essential to identify the local probability of platelet capture.
Fig.~\ref{Figure_3}~A,B show representative platelet trajectories at $\textit{Ht} = 0.33$ and $\textit{Ht} = 0.48$ in the $y$-$z$ plane of the simulated system, over $200~t_\text{adv}$.
The solid black lines denote the walls (straight upper wall and wavy lower wall), while the dashed line represents the capture distance.
Coloured trajectories highlight platelets that cross the capture distance at least once, whereas grey trajectories correspond to platelets that never reach it during the simulation window.
Fig.~\ref{Figure_3}~C,~D provide complementary representations of platelet-wall distance over time, expressed in units of $t_\text{adv}$.
Purple lines denote platelets initialised in the upper half of the channel ($z/L_z > 0.5$), while green lines denote those initialised in the lower half ($z/L_z \leq 0.5$).

From these panels, it can be observed that only a limited fraction of platelets, primarily those already located near the wall (within $\approx~0.2~L_z$), are able to marginate and reach the capture distance within the simulated time frame.
This observation indicates that the simulations are relatively short compared to the total time scale over which platelet margination occurs.

The probability of platelet capture was quantified as the net number of platelet transitions into the region within capture distance (dashed lines at $1.5~r_{\text{plt}}$ from the wall; Fig.~\ref{Figure_3}).
Each transition toward the wall (capture) was counted positively, while transitions back into the bulk (escape) were counted negatively.
The net number of transitions was then normalised per sinusoidal wave and per 100 platelets in a simulated volume of $\sim 5.7 \times 10^{4}\,\mu\mathrm{m}^{3}$.

\begin{figure*}[htb]
 \centering
 \includegraphics[width=0.8\textwidth]{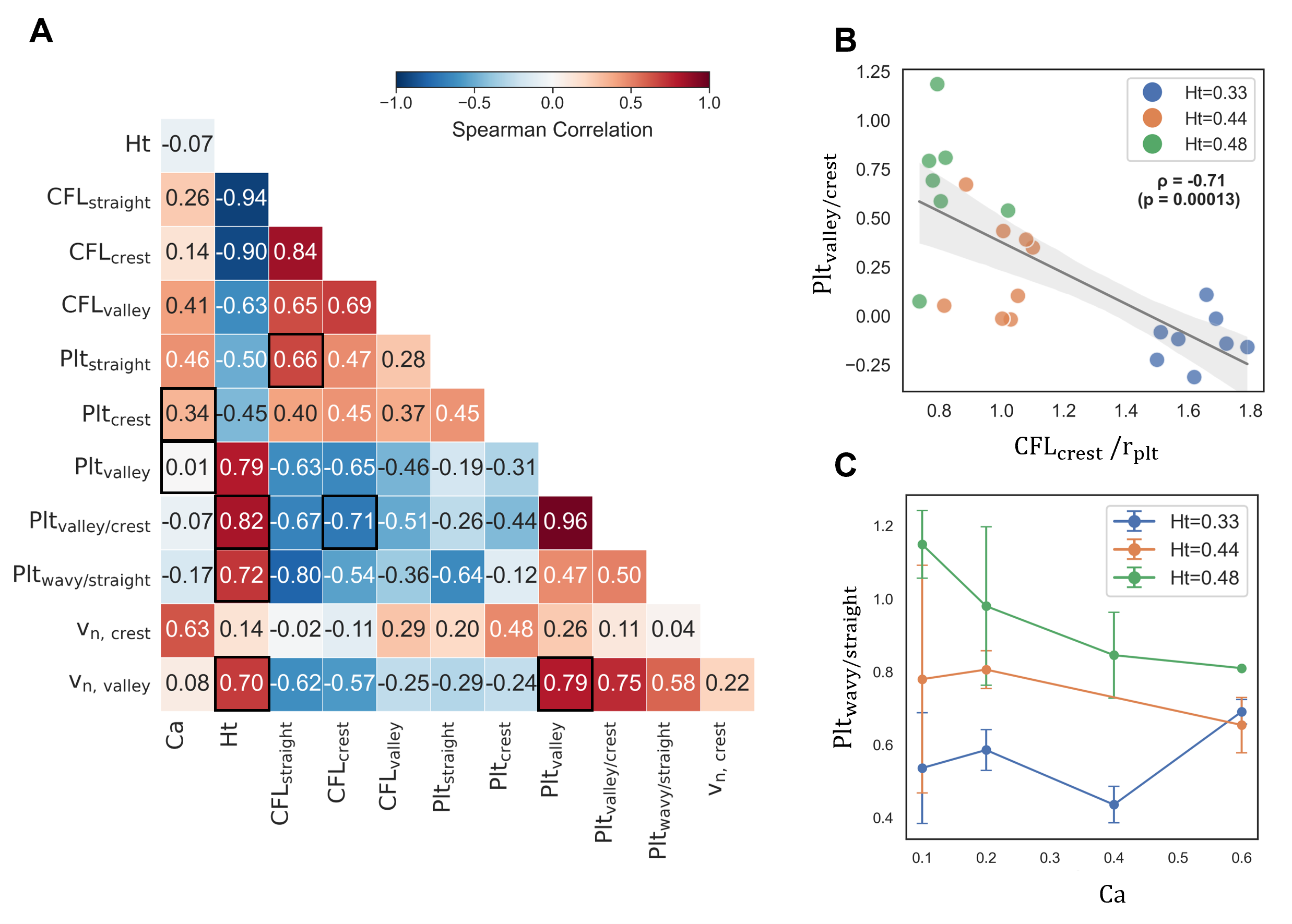}
 \caption[Correlations between platelet capture, near-wall platelet velocity, CFL and wall shear rate]{Correlations between platelet capture, near-wall platelet velocity, CFL and wall shear rate. \textbf{(A)} Spearman correlation matrix including: \textit{Ca}, \textit{Ht}, CFL thickness at the straight upper wall ($\text{CFL}_{\text{straight}}$), at the crest ($\text{CFL}_{\text{crest}}$) and at the valley ($\text{CFL}_{\text{valley}}$) of the wavy wall; probability of platelet capture at the straight wall ($\text{Plt}_{\text{straight}}$), at the crest ($\text{Plt}_{\text{crest}}$) and at the valley ($\text{Plt}_{\text{valley}}$) of the wavy wall; and at the wavy lower wall ($\text{Plt}_{\text{wavy}}$); platelet capture ratio between valley and crest ($\text{Plt}_{\text{valley}} / \text{Plt}_{\text{crest}}$), and between wavy and straight wall ($\text{Plt}_{\text{wavy}} / \text{Plt}_{\text{straight}}$); mean normal velocity of platelets within a $1.5$--$2.5~r_{\text{plt}}$ band from the wall at the crest ($v_{n,\text{crest}}$) and valley ($v_{n,\text{valley}}$). Black squares highlight the correlations specifically discussed in the main text. \textbf{(B)} Scatter plot of $\text{Plt}_{\text{valley}} / \text{Plt}_{\text{crest}}$ as a function of $\text{CFL}_{\text{crest}} / r_{\text{plt}}$ at three \textit{Ht} values ($Ht = 0.33$, blue; $0.44$, orange; $0.48$, green). The black line represents the linear regression fit, while the grey shaded area indicates its 95\% confidence interval. The Spearman correlation coefficient ($\rho$) and the corresponding FDR-adjusted $p$-value are reported. \textbf{(C)} $\text{Plt}_{\text{wavy}} / \text{Plt}_{\text{straight}}$ as a function of \textit{Ca} for three \textit{Ht} values ($Ht = 0.33$, blue; $0.44$, orange; $0.48$, green). Means $\pm$ standard deviations from at least two independent simulations per condition are shown. Lines are guides for the eye.}
 \label{Figure_5}
\end{figure*}

Fig.~\ref{Figure_4} displays the resulting spatial distribution of platelet capture probability along half of the wave at $\textit{Ht} = 0.33$, $0.44$ and $0.48$.
The platelet capture probability is averaged over different simulations at varying \textit{Ca}, since platelet capture at the crest and valley was found to be largely insensitive to \textit{Ca} (crest: $\rho = 0.34$, $p = 0.11$; valley: $\rho = 0.01$, $p = 0.98$; Fig.~\ref{Figure_4}~A).
This finding agrees with previous studies showing that the shear-rate dependence of particle margination is most pronounced at low shear rates~\cite{Mueller2014}; 
whereas in the tank-treading regime of RBCs ($\textit{Ca} \gtrsim 0.2$), platelet margination becomes primarily modulated by \textit{Ht}~\cite{Spann2016} and confinement~\cite{Krueger2015}.

In Fig.~\ref{Figure_4}, positive values (red regions) of the net number of transitions across the capture distance indicate preferential platelet entry and accumulation in the near-wall region, whereas negative values (blue regions) denote the tendency to escape toward the bulk.
At low \textit{Ht} ($Ht= 0.33$), platelets preferentially accumulate near the crest, while they tend to escape from the valley.
At intermediate \textit{Ht} ($Ht = 0.44$), platelet capture becomes positive throughout the sinusoid, with a stronger accumulation at the crest.
At higher \textit{Ht} ($Ht = 0.48$), the capture probability becomes more uniform along the entire sinusoidal wave.

This preferential platelet distribution agrees with the mean platelet velocity in the near-wall region, shown by arrows in Fig.~\ref{Figure_4}.
Velocities are consistently higher near the crest than in the valley.
The wall-normal component of the platelet velocity near the valley increases with \textit{Ht} ($\rho = 0.70$, $p = 2 \times 10^{-4}$; Fig.~\ref{Figure_5}~A) and is strongly correlated with platelet accumulation in this region ($\rho = 0.79$, $p = 7.5 \times 10^{-6}$; Fig.~\ref{Figure_5}~A).

Overall, these results indicate that increasing \textit{Ht} promotes a more uniform platelet capture along the wavy wall.
Interpreting the sinusoidal wave as an arrangement of platelet aggregates aligned with the flow direction (Fig.~\ref{Figure_1}), our results suggest that \textit{Ht} governs the morphological evolution of such aggregates over time.
At low \textit{Ht}, capture occurs predominantly at crests, favouring the development of sharper, high-amplitude aggregates.
At higher \textit{Ht}, capture becomes more evenly distributed, leading to aggregates with a less pronounced amplitude.
This trend is consistent with experimental observations \cite{Pero2024}, showing increased platelet adhesion~\cite{Chen2013} and surface coverage at higher \textit{Ht}~\cite{Pero2024, Spann2016}.

\begin{figure*}[htb]
 \centering
 \includegraphics[width=0.7\textwidth]{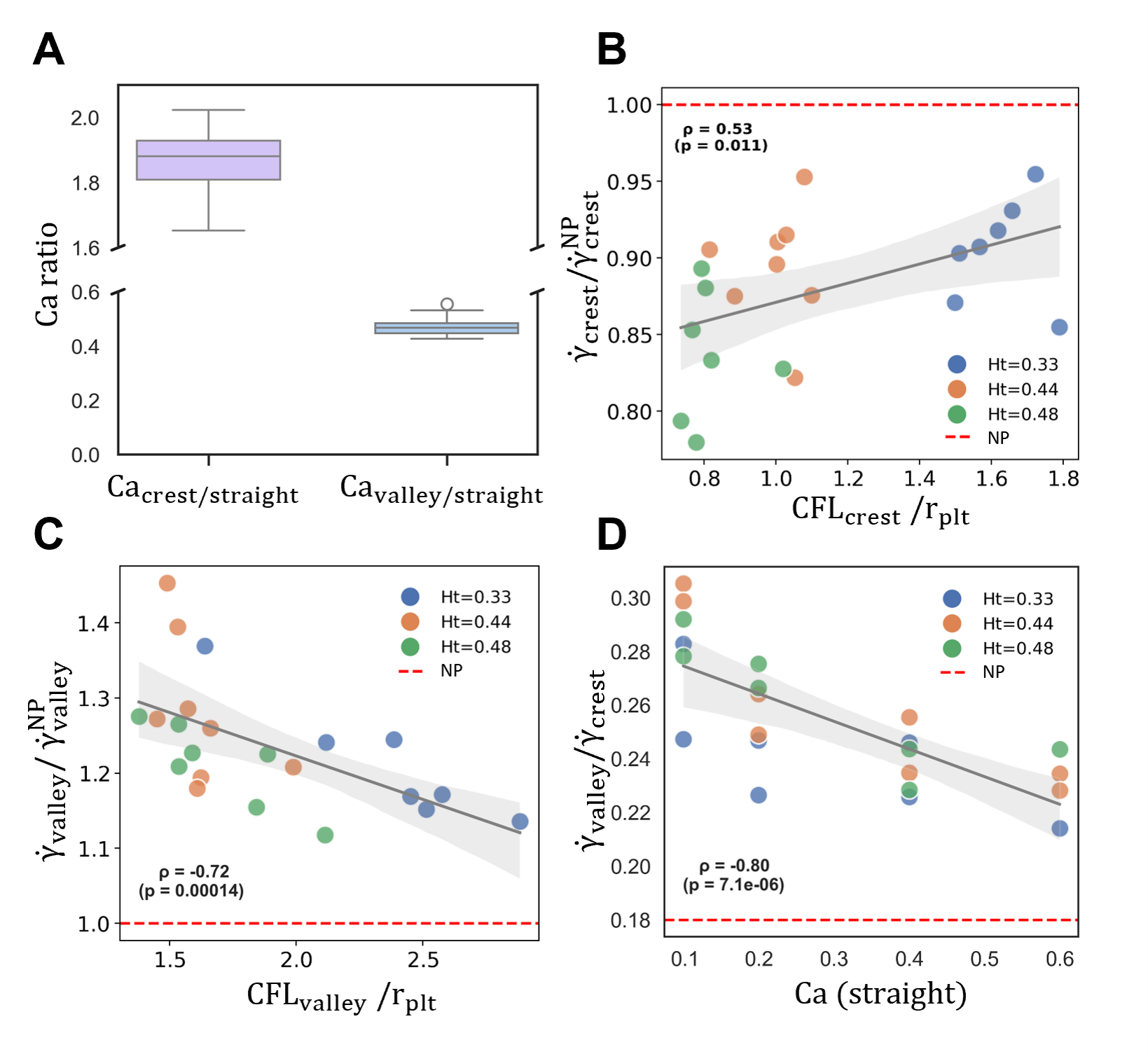}
 \caption{Crest-valley asymmetry in wall shear rate shaped by CFL and RBC deformability. Wall shear rate was quantified from velocity profiles by applying bilinear interpolation to the lattice velocity data followed by parabolic fitting. The resulting values were averaged spatially along the flow direction, temporally over the last quarter of the simulation ($150-200~t_{\text{adv}}$), and across the different sinusoidal periods of the computational domain. Additional averages were obtained over independent simulations with randomised initial particle positions. \textbf{(A)} Box plot includes all simulations across \textit{Ca} and \textit{Ht} conditions. The purple box (left) indicates the ratio of \textit{Ca} at the crest to the straight wall, while the blue box (right) indicates the ratio at the valley. An axis break between $0.6$ and $1.6$ improves visibility. \textbf{(B--C)} Shear rate (normalised to the no-particle reference) at the crest (B) and valley (C) as a function of local CFL thickness normalised by platelet radius. \textbf{(D)} Valley-to-crest ratio as a function of \textit{Ca} (at the straight wall), for $Ht = 0.33$ (blue), $0.44$ (orange), and $0.48$ (green). The red dashed line marks the particle-free reference case. Black lines show linear regression fits, with grey bands indicating the $95\%$ confidence interval. Reported are Spearman correlation coefficients ($\rho$) and FDR-adjusted $p$-values.}
 \label{Figure_6}
\end{figure*}

The thickness of the CFL near the wall plays a central role in determining the degree of platelet margination.
Spearman correlations between \textit{Ca}, \textit{Ht}, CFL, and the probability of platelet capture at the straight wall, crest, and valley are reported in the matrix in Fig.~\ref{Figure_5}~A.
Platelet accumulation at the straight upper wall is positively correlated with the CFL at the same location ($\rho = 0.66$, $p = 0.005$).
In our simulations, platelets have a diameter of $2~\si{\micro\meter}$, while the CFL at the straight wall ranges from $1.0$ to $2.4~\si{\micro\meter}$.
This trend supports the concept, previously highlighted by Müller~\textit{et al.}~\cite{Mueller2014, Mueller2016}, that particles are more likely to marginate when the CFL thickness becomes comparable to their size.

In the presence of a wavy wall, the suspension dynamics becomes more complex.
Spatial variations of the CFL along the sinusoidal profile make crests and valleys effectively compete for platelet capture near the lower wall.
At the crest, the CFL ranges from $0.7$ to $1.8~\si{\micro\meter}$, comparable to or smaller than the platelet diameter, whereas at the valley it spans $1.4$ to $2.9~\si{\micro\meter}$, approaching or exceeding platelet size.
As \textit{Ht} increases, the CFL becomes thinner at both locations, and platelet capture at the valley progressively increases until reaching levels comparable to the crest ($\text{Plt}_{\text{valley}} / \text{Plt}_{\text{crest}} \approx 1$, Fig.~\ref{Figure_5}~B).

Among these effects, the reduction of the CFL at the crest emerges as the factor most closely associated with platelet redistribution toward the valley, as suggested by its strongest correlation with $\text{Plt}_{\text{valley}} / \text{Plt}_{\text{crest}}$ ($\rho = -0.71$, $p = 0.0013$).
Since platelets tend to marginate most effectively where the CFL is comparable to their size~\cite{Mueller2014, Mueller2016}, a narrowing of the CFL at the crest below platelet diameter may render this region less favourable, with platelets increasingly directed toward the valley, where the CFL becomes closer to their size.
Additionally, platelet accumulation is generally lower on the wavy wall than on the straight wall, as reflected by the ratio $\text{Plt}_{\text{wavy}} / \text{Plt}_{\text{straight}}$ being mostly below unity (Fig.~\ref{Figure_5}~C).
This ratio approaches unity at low \textit{Ca} and high \textit{Ht}, but levels off around $0.7$ at higher \textit{Ca}.

\subsection{Crest–valley shear rate gradient and its potential implications for platelet adhesion mechanisms}
\label{sec:Res_WSR}

The wall shear rate ($\dot{\gamma}_{\text{w}}$) is a key factor in platelet adhesion.
In low-shear regions, such as cavities or stagnant zones, adhesion is primarily mediated by collagen, fibrinogen, or fibronectin~\cite{Savage1996}, and is favoured by the longer residence time and reduced washout~\cite{Herbig2017}.

At shear rates above $\sim 500~\si{\per\second}$, fibrinogen-mediated adhesion becomes less efficient, and platelet tethering requires immobilised von Willebrand factor (vWF)~\cite{Jackson2007}.
At even higher shear rates ($>1000~\si{\per\second}$), shear forces induce conformational changes in vWF that expose A1 binding sites for platelet GPIb$\alpha$~\cite{Schneider2007}.
In our simulations, we set \textit{Ca} based on the shear rate at the straight upper wall, with values ranging from $0.1$ to $0.6$.
This corresponds to straight-wall shear rates of $133$--$795~\si{\per\second}$, thereby covering different adhesion regimes.

Box plots in Fig.~\ref{Figure_6}~A include all simulations across the different \textit{Ca} and \textit{Ht} conditions.
They show that the effective \textit{Ca} at the crest is about twice the value at the straight wall, with maximum values exceeding the $1000~\si{\per\second}$ threshold where platelet adhesion is mediated by vWF unfolding.
In contrast, near the valley, \textit{Ca} is roughly half the value of the straight wall, indicating that adhesion in that region is more likely governed by low-shear mechanisms mediated by collagen/fibrinogen.

Fig.~\ref{Figure_6}~B,~C report the shear rate at the crest and valley (normalised with respect to the particle-free reference) as a function of local CFL thickness, while Fig.~\ref{Figure_6}~D shows the valley-to-crest ratio as a function of \textit{Ca} at the straight wall.
The red-dashed line represents the no-particle (NP) reference.
These results highlight the interplay between RBC proximity to the wall (CFL) and their deformability (\textit{Ca}) in shaping the shear profile along the sinusoidal wall.

In the absence of particles, the shear rate near the valley is about $18\%$ of that near the crest (Fig.~\ref{Figure_6}~D).
The introduction of RBCs perturbs this balance: reduced CFL thickness correlates with lower shear rate at the crest (Fig.~\ref{Figure_6}~B) and increased shear rate at the valley 
(Fig.~\ref{Figure_6}~C).
We hypothesise that RBCs introduce flow resistance at the crest, reducing $\dot{\gamma}_{\text{crest}}$, while enhancing velocity gradients in the otherwise stagnant valley region, thereby increasing $\dot{\gamma}_{\text{valley}}$.
Moreover, the valley-to-crest ratio increases with decreasing \textit{Ca} (\textit{i.e.}, with less deformable RBCs), reaching up to $28\%$ (Fig.~\ref{Figure_6}~D). 
This finding supports the view that, at low \textit{Ca}, reduced RBC deformation and flow adaptation may amplify flow disturbances, leading to greater deviations in the near-wall shear distribution compared to the particle-free reference.

In summary, RBC proximity to the wall (reduced CFL), combined with a limited deformability at low \textit{Ca}, tends to slightly reduce the wall shear rate gradient along the sinusoidal surface.
Nevertheless, within the range of \textit{Ca} and \textit{Ht} explored, substantial gradients remain, with the shear rate near the valley reaching at most $\sim 28\%$ of that near the crest (Fig.~\ref{Figure_6}~A).

\section{Conclusion}
\label{sec:conclusions}

The geometry of blood vessels plays a crucial role in hemostatic and thrombotic processes by regulating the near-wall dynamics of red blood cells (RBCs) and platelet margination. At the same time, growing platelet aggregates continuously reshape the vessel wall topography, leading to a strongly coupled system. 
However, how such surface heterogeneities influence local hemodynamics and thereby regulate the spatial growth of platelet aggregates remains poorly understood.
To address this gap, we simulated blood flow in a straight channel with a sinusoidal bottom wall as a model for platelet aggregates already adhered to the bottom wall.
Blood is modelled as a suspension of resolved red blood cells (RBCs) and platelets using a combination of the lattice Boltzmann, immersed boundary, and finite element methods.
The aim of this study was to investigate how the sinusoidal geometry modulates the local cell-free-layer (CFL), platelet margination
and wall shear rate, at different values of haematocrit ($Ht = 0.33$, $0.44$ and $0.48$) and capillary number ($\textit{Ca} = 0.1$, $0.2$, $0.4$ and $0.6$).

A key result is that the CFL thickness is non-uniform along the sinusoidal surface and tends to thicken in the valley, particularly at high \textit{Ca} and low \textit{Ht}.
Under these conditions, RBCs tend to organise into ordered and packed structures, possibly explaining the increased RBC wall lift and the thicker CFL in the valley.

Furthermore, we show that platelets preferentially marginate where the CFL thickness matches the platelet size, in accordance with literature findings~\cite{Mueller2014, Mueller2016}.
The local CFL thickness, in turn, depends on the average \textit{Ht}. This mechanism drives a competition between crests and valleys. At low \textit{Ht}, platelets preferentially accumulate near the crest, where the CFL thickness is of the order of the platelet size.
This preferential accumulation at the crest is thought to promote the growth of thicker and high-amplitude aggregates.
Contrarily, at high \textit{Ht}, the CFL near the crest decreases markedly, driving platelet margination toward the valley.
This effect results in a more homogeneous accumulation along the sinusoidal surface, in agreement with experimental observations~\cite{Pero2024, Colace2010}.

Although platelet adhesion is not explicitly modelled, the computed shear landscape helps identify distinct adhesion pathways.
The wall shear rate along the sinusoidal bottom wall exhibits a pronounced gradient between the crest and the valley, spanning shear regimes that are known to activate different platelet adhesion mechanisms. While shear rates at the crest reach levels compatible with vWF-mediated GPIb$\alpha$ interactions, valley regions remain in a lower-shear regime more consistent with collagen- or fibrinogen-mediated adhesion. 

Overall, these findings provide a mechanistic interpretation of platelet aggregate growth patterns by linking preferential sites of platelet accumulation to local variations in the CFL thickness and wall shear rate.
Looking forward, identifying the most probable adhesion sites together with their local shear environment provides a rational basis to optimise drug delivery systems that exploit flow-mediated targeting~\cite{ZeibiShirejini2023, Molloy2017}.
Moreover, the ability to hypothesize the molecular players involved in adhesion at specific sites points to candidate targets for pharmacological modulation.
Therefore, this work contributes not only to a better understanding of thrombus morphology, but also to the design of precision antithrombotic therapies and targeted drug delivery approaches~\cite{Barba2017}.

\section*{Acknowledgements}
The authors thank Benjamin Owen, Fatemehsadat Mirghaderi, and Roslyn Hay for helpful discussions during the development of this work.
\section*{Author contributions: CRediT}
EP: conceptualization, investigation, data curation, formal analysis, visualization, writing - original draft; GT and SG: conceptualization, supervision, writing-review; CD: methodology, writing-review;  TK: conceptualization, methodology, software, resources, supervision, writing-review and editing. 
\section*{Funding sources}
This work was supported by the University of Naples "Federico II" (PhD scholarship and mobility founding, project code 000008-DOTTORATO-38-CICLO-DICMAPI). Financial support from the 2022 University Research Funding Program of the University of Naples Federico II for the project “SHEAR – Solid Handling and Efficient Mixing at the Micro-scale: A Spatio-Temporal Analysis in Microfluidic Reactors", E63C22002450005, is gratefully acknowledged.
\section*{Declaration of Interest}
The authors declare that they have no conflicts of interest.

\bibliographystyle{unsrt}  
\bibliography{references}  

\end{document}